\documentclass[12pt]{article}
\usepackage{epsfig,amsmath,amssymb,euscript}
\usepackage[square,comma,sort&compress,numbers]{natbib}
\usepackage{color}
\usepackage{graphicx}
\usepackage{lineno}

\begin{document}
\title{Rising Above Chaotic Likelihoods}


\author{Hailiang Du$^{1,2}$ \quad \quad  Leonard A. Smith$^{1,2,3}$ \\
        $^1$Centre for the Analysis of Time Series,\\
       London School of Economics, London WC2A 2AE. UK\\
       $^2$Center for Robust Decision Making on Climate and Energy Policy,\\
       University of Chicago, Chicago, IL 60637, US\\
   $^3$Pembroke College, Oxford, UK}

\date{\today}

\maketitle

\begin{abstract}

Berliner (Likelihood and Bayesian prediction for chaotic systems, J. Am. Stat. Assoc. 1991) identified a number of difficulties in using the likelihood function within the Bayesian paradigm which arise both for state estimation and for parameter estimation of chaotic systems. Even when the equations of the system are given, he demonstrated ``chaotic likelihood functions" both of initial conditions and of parameter values in the Logistic Map. Chaotic likelihood functions, while ultimately smooth, have such complicated small scale structure as to cast doubt on the possibility of identifying high likelihood states in practice. In this paper, the challenge of chaotic likelihoods is overcome by embedding the observations in a higher dimensional sequence-space; this allows good state estimation with finite computational power. An importance sampling approach is introduced, where Pseudo-orbit Data Assimilation is employed in the sequence-space, first to identify relevant pseudo-orbits and then relevant trajectories. Estimates are identified with likelihoods orders of magnitude higher than those previously identified in the examples given by Berliner. {Pseudo-orbit Data Assimilation importance sampler exploits the information both from the model dynamics and from the observations. While sampling from} the relevant prior {(here, the natural measure)} will, of course, eventually yield an accountable sample, given the realistic computational resource this traditional approach would provide no high likelihood points at all. {While one of the challenges Berliner posed is overcome, his central conclusion is supported.} ``Chaotic likelihood functions" for parameter estimation still pose a challenge; this fact {helps clarify} why physical scientists maintain a strong distinction between the initial condition uncertainty and parameter uncertainty.

\end{abstract}

\section{Introduction}

Nonlinear chaotic systems pose several challenges both for state estimation and for parameter estimation. Chaos as a phenomenon implies sensitive dependence {to} initial condition: initially nearby states will eventually diverge in the future. The bifurcations of various chaotic systems~\cite{Sprott2003} reveal how the behavior of the system differs as a parameter value changes. One might think that Bayesian analysis should be able to obtain good estimation both of initial conditions and of parameter values without much trouble. Berliner~\cite{Berliner1991} examined the log-likelihood function of estimates of initial conditions and parameter values for the Logistic Map{, noting} that chaotic systems can lead to ``chaotic likelihood functions'' {and} suggesting that Bayesian analysis would require prohibitively intensive computing. The failure of variational approaches, when applied to long window observations of chaotic systems~\cite{Miller1994,Judd2001,DuPDA}, supports his point. Sensitivity to initial condition on the other hand suggests that information in the observations (even over a relatively short window) can lead to good estimates of the initial condition~\cite{Judd2004}. An importance sampling approach to Berliner's challenge without ``intensive computing'' is deployed in this paper. Adopting Pseudo-orbit Data Assimilation (PDA)~\cite{Judd2001,DuPDA} recasts the {problem} into a higher dimensional sequence space, where truly high likelihood states are successfully located. 
The challenges of initial condition estimation and parameter estimation are dissimilar for chaotic systems. PDA does not easily generalize to parameter estimation, as it is unclear how to mathematically define a relevant subspace of {the} parameter space in which {the} high likelihood trajectories might exist. Thus challenges remain in identifying high likelihood parameter values given the initial condition; this asymmetry reflects differences between the initial conditions and parameter values. In terms of estimating initial conditions given the parameter values, however, Berliner's challenge as originally stated is met and resolved.


\section{Chaotic Likelihood Function of Initial Conditions}

{Berliner's \cite{Berliner1991} formulation of the problem is adapted here. The Logistic Map is the system{; in sections 2-4, }the parameter $a=4$ is known but the true initial state $\tilde{x}_{0}$ is not. }In that case, the experiment is said to fall within the perfect model scenario\footnote{{The perfect model scenario, when the system and the model are identical, eases use of a digital computer; one avoids the issue of ``round-off error'' by consistently use of the same computer and code. This implies, of course, that Equation~\ref{eq:model} no longer reflects the perfect model. }For discussion, see~\cite{Smith87, Beck92, Smith02} and references thereof.}. In the first four sections of this paper, the mathematical system and the model are identical, the word model is not used. Once real systems are considered, either structural model error or parameter inaccuracy requires one to distinguish the system which generated the data from the model(s) employed to analyze it.

The evolution of system states\footnote{For $m = 1$, the state $x_i$ is a scalar.} $x_{i}\in \mathbb{R}^m$ is then governed by the nonlinear dynamics $f:x_{i+1}=f(x_{i})$, where for the Logistic Map
\begin{eqnarray}
 \label{eq:model}
 f(x_{i})=ax_{i}(1-x_{i}).
\end{eqnarray}
Assuming additive observational noise, $\delta_{i}$, yields observations, $s_{i}=\tilde{x}_{i}+\delta_{i}$ where $\tilde{x}$ is the true system state (Truth) and the observational noise, $\delta_{i}$, is Independent Normally Distributed (IND), $\delta_{i}\sim N(0,\sigma^{2})$. Under this assumption of normality, the log-likelihood function is:
\begin{eqnarray}
 \label{eq:LLik}
     LLik(x_{0})=-\sum_{i=1}^{n-1}(s_{i}-f^{i}(x_{0}))^2/2\sigma^{2},
\end{eqnarray}
where $f^{i}(x)$ is the $i^{th}$ iteration of $f(x)$, $s_{i}$ is the $i^{th}$ observation, and $n$ is the duration of observations considered.

\begin{figure}[h]
  \hbox{
  \epsfig{file=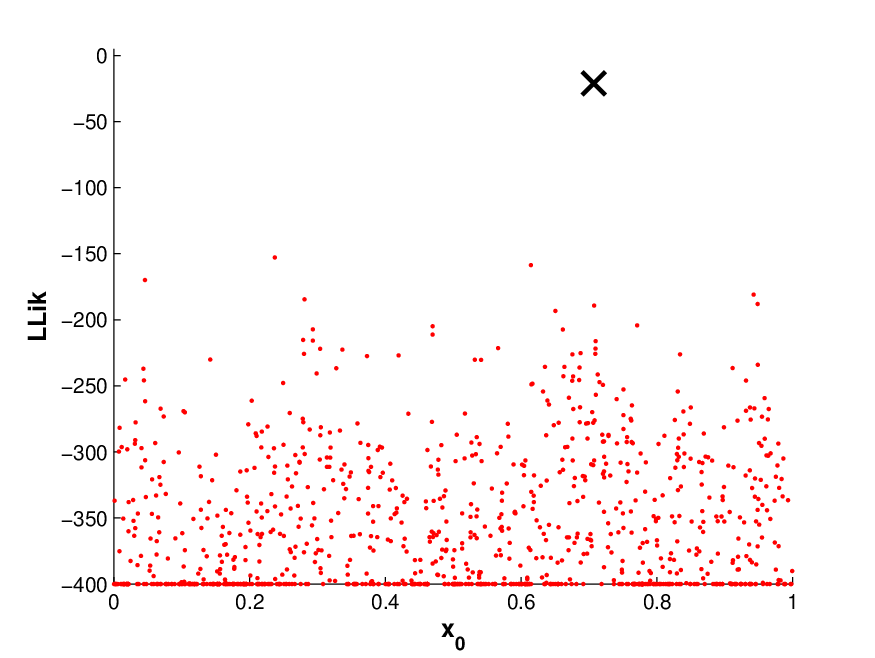, width=7cm}
  \epsfig{file=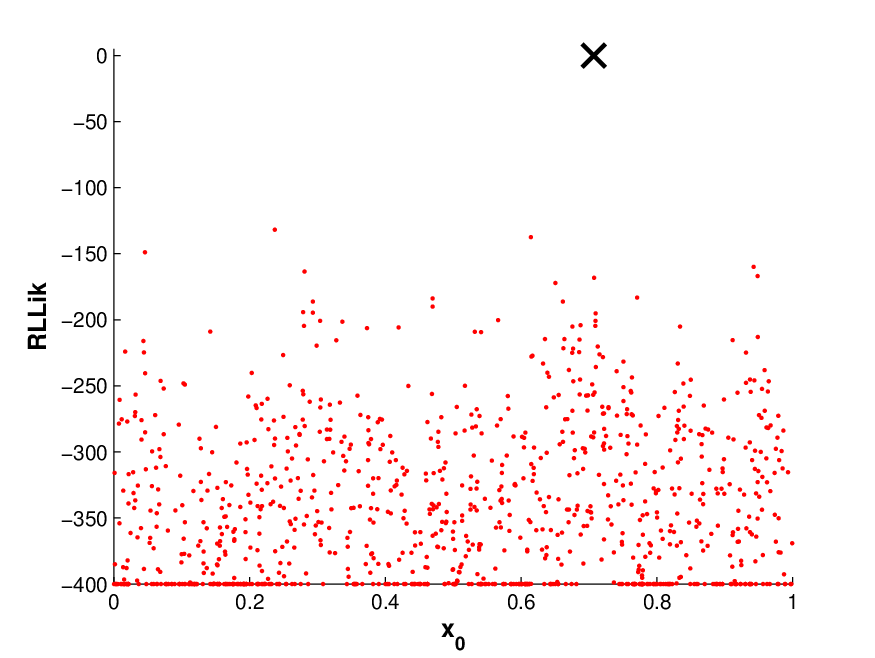, width=7cm}
}

\caption{Typical log-likelihood of 1024 states (uniformly distributed on $[0,1]$) for the Logistic Map. The true initial condition $\tilde{x}_{0}=\sqrt{2}/2$ {(denoted by `$\times$')}, $\sigma=0.1$ and $n=32$. a) Log-likelihood function, b) Relative log-likelihood to $\tilde{x}_{0}$. States which have (relative) log-likelihood less than -400 are plotted on the -400 horizontal line. All logarithms are using natural base. {The maximum likelihood value found was $\sim e^{-109}$, which corresponds to log-likelihood of $-132$ relative to Truth.}}
  \label{fig:LLikNormal}
\end{figure}

Figure~\ref{fig:LLikNormal} shows the chaotic likelihood structure of 1024 {candidate values of $x_{0}$ drawn uniformly on the interval zero, one.} Panel (a) plots the log-likelihood for {each} $x_{0}$, this can be contrasted with various panels in Berliner's~\cite{Berliner1991} Figure 3.\footnote{Here log-likelihood functions based on a sequence of 32 observations are computed because problem becomes more obvious when more observations are used. Berliner examined 15 \& 10 observations. Shorter sequences of observations are examined in Section 3.} Panel (b) shows the log-likelihood relative to that of the true trajectory of the system state\footnote{Note: given that only finite observations are considered the true state of the system is, with probability 1, not the maximum likelihood state.}. For the convenience of illustration, the same normalization used in panel (b) is applied in Figures 2-4 in this paper. From Figure~\ref{fig:LLikNormal}, it is clear that no high likelihood states are identified. This is not a case of equifinality\footnote{Equifinality occurs when many potential solutions to a task are each good, making it impossible to identify the true solution given the information in hand. {In such cases the sampled likelihood functions are} relatively flat. Equidismality arises when the sampled relative likelihood function is flat yet all solutions tested have vanishingly small likelihood given the information. Examining the relative likelihood obscures the difference; fortunately the expected (distribution of) likelihood can be computed from the noise model alone without knowledge of the true initial condition.}.

\begin{figure}[h]
  \hbox{
  \epsfig{file=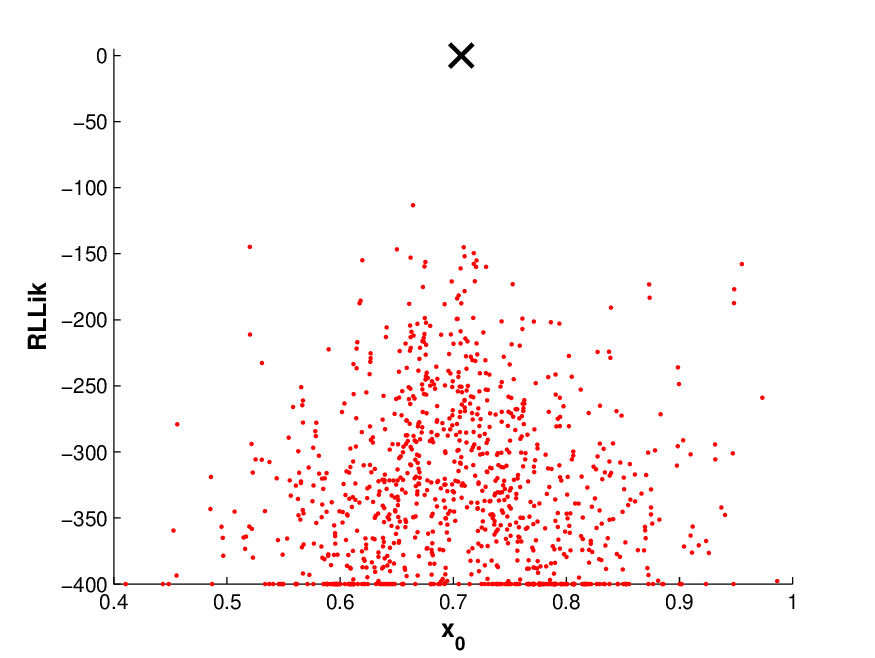, width=7cm}
  \epsfig{file=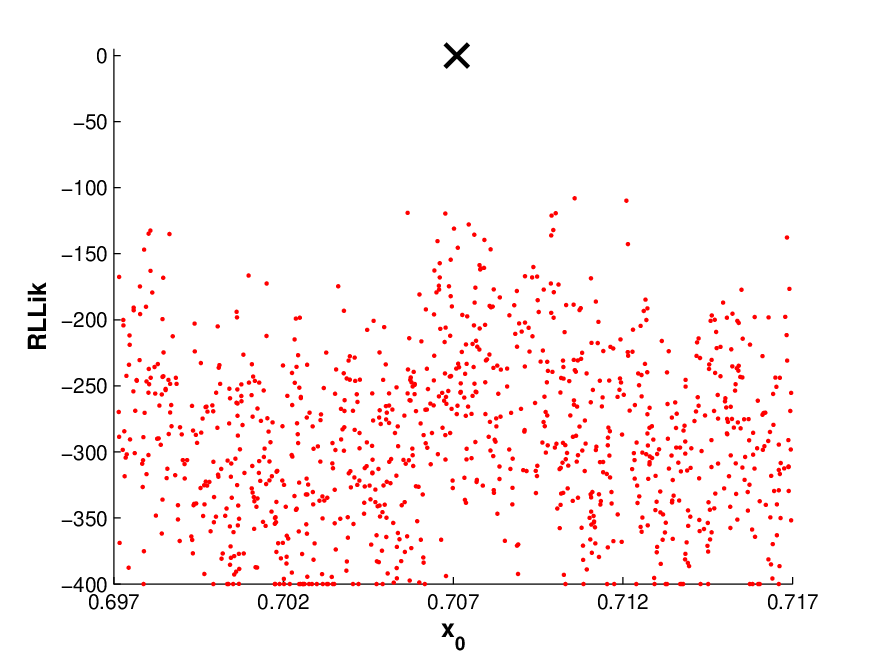, width=7cm}
}

\caption{{Relative log-likelihood of 1024 states a) sampled from inverse observational noise, b) uniformly sampled from $[\tilde{x}_{0}-\frac{\sigma}{10},\tilde{x}_{0}+\frac{\sigma}{10}]$. The true initial condition $\tilde{x}_{0}=\sqrt{2}/2$ (denoted by `$\times$'), $\sigma=0.1$ and $n=32$.}}
  \label{fig:LLikZoom}
\end{figure}

Given the observational noise distribution, one can add random draws from the inverse of the observational noise distribution to the observation to obtain candidate estimates of initial condition. Figure~\ref{fig:LLikZoom}a shows the relative log-likelihood of 1024 samples from inverse observational noise. No high likelihood states are identified in this way. {(The maximum likelihood value found was $\sim e^{-90}$.)} To illustrate the impact of making much more precise observations, consider a case where $\tilde{x}_{0}$ is known to be within a region of radius only $\sigma/10$. Figure~\ref{fig:LLikZoom}b shows the relative log-likelihood of 1024 uniformly sampled states in the region around Truth with $\sigma/10$ radius. Yet again, no high likelihood state are identified. {(The maximum likelihood value found increases to $\sim e^{-85}$.)}

\begin{figure}[h]
\centering
  \epsfig{file=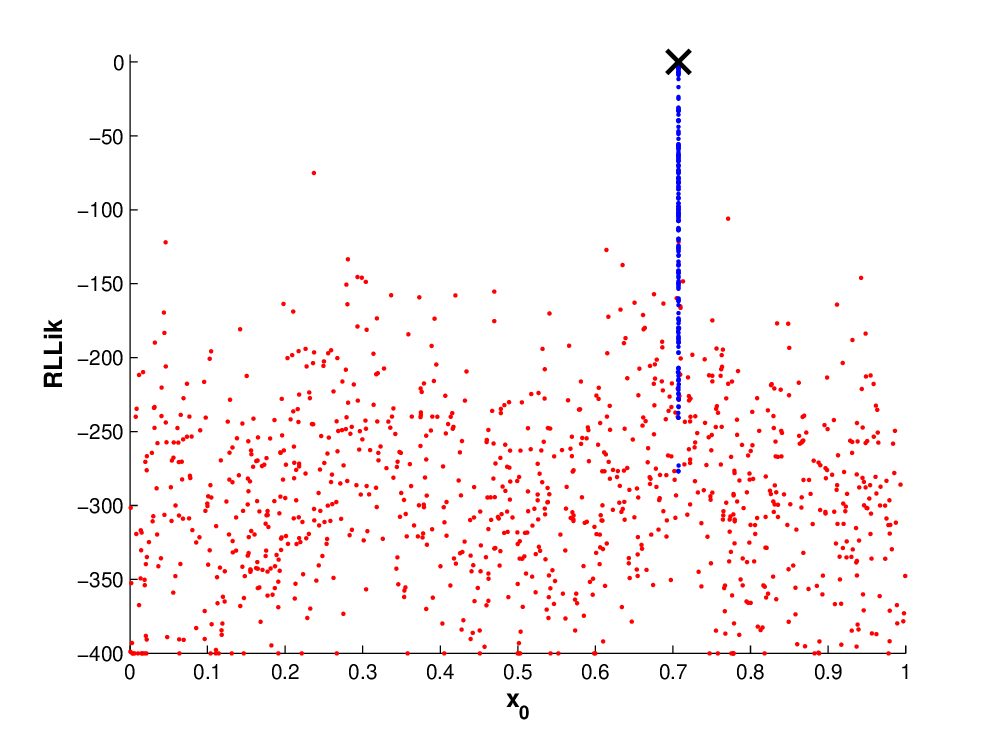, width=7cm}
  \epsfig{file=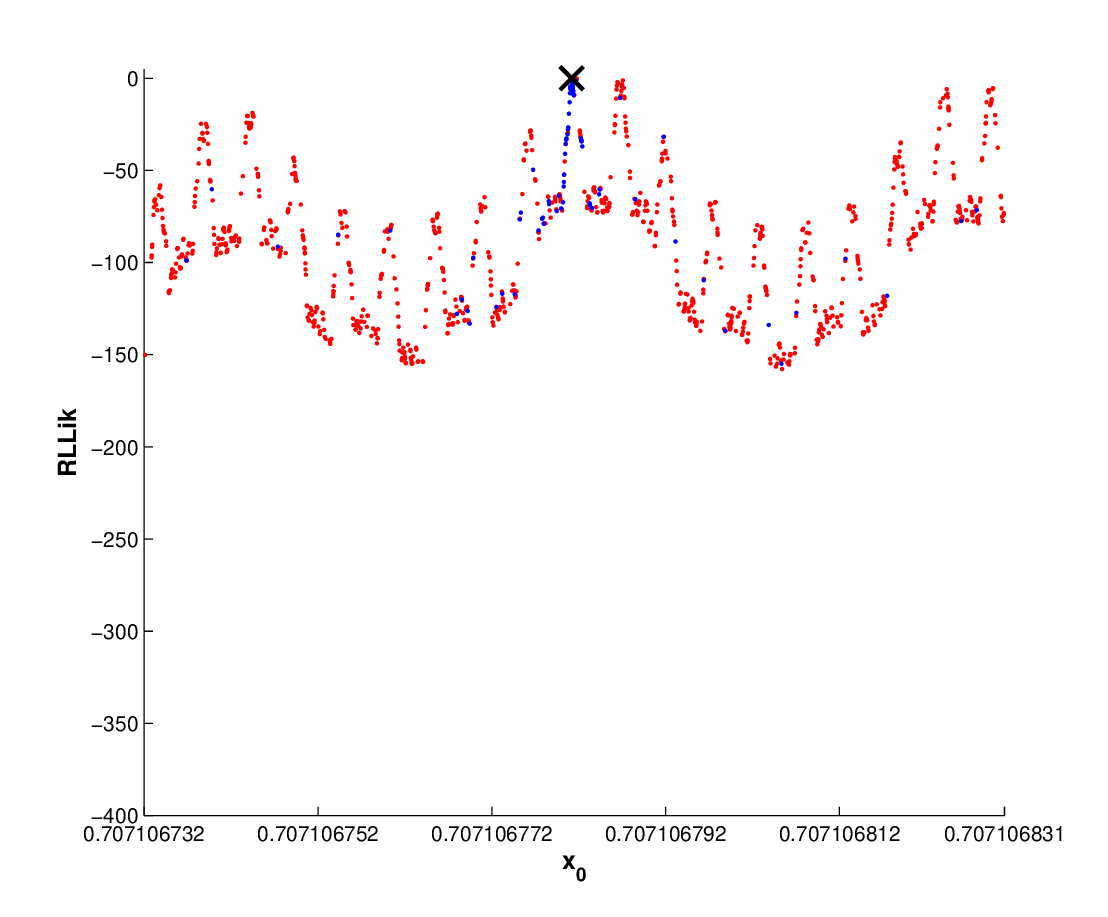, width=7cm}

\caption{a) Following Figure~\ref{fig:LLikNormal}b, {{but in this case} including a set of 1024 states (blue), }which are extremely close to $\tilde{x}_{0}$, generated by spiral sampling around the $\tilde{x}_{0}$ {(see footnote 6)}; b) zoom in of a).}
  \label{fig:LLikBlue}
\end{figure}
This difficulty here { is not a shortcoming of the likelihood framework,} as there are high likelihood states other than Truth. One may demonstrate that such high likelihood states exist by sampling the points on a logarithmic spiral approaching Truth (to machine precision)\footnote{In this experiments 1024 points are generated by $\tilde{x}_{0}+2^{-(10+\frac{60i}{1024})}\epsilon_{i}, i=1,2,...,1024$ where $\epsilon_{i}$ is random drawn from $U(0,1)$.}. Figure~\ref{fig:LLikBlue} {demonstrates} { that high likelihood states other than Truth {do }exist (i.e. some of the blue points).} A smooth curve of the log-likelihood function is only observed within a radius of $\tilde{x}_{0}$ smaller than $\sim 10^{-7}$, {(see Figure~\ref{fig:LLikBlue}b)}.

Without knowing Truth, of course, this approach to identifying {high likelihood} points is inaccessible. The likelihood function is extremely jagged; as Berliner~\cite{Berliner1991} stressed, finding even one high likelihood state by sampling the state space is prohibitively costly.  That said, there is no sense in which ``sensitivity to the initial conditions'' can be taken to imply that the information in the initial condition is ``forgotten'' or ``lost". There is sufficient information in the observation segment to {distinguish} high likelihood initial states.
Candidate states with non vanishing log-likelihood relative to Truth can be found by extracting the information from the system dynamics using a relatively new approach to data assimilation: Pseudo-orbit Data Assimilation.

\section{Importance Sampling via Pseudo-orbit Data Assimilation }

\subsection{Methodology}

{Uniform sampling in state space is not an efficient approach to locating high likelihood states.} As the dimension of the system increases, 
the task {becomes even more} computationally impractical. PDA importance sampler\footnote{In high-dimensional space the sampler targets the relevant lower-dimensional trajectory manifold which is more efficient than sampling a hypersphere suggested by one observation. Even in the one{-}dimensional Logistic Map this approach succeeds by using PDA to sample the trajectory manifold in the n-dimensional sequence space.} locates high likelihood states in the trajectory manifold by adopting the Pseudo-orbit Data Assimilation approach~\cite{Judd2001,DuPDA}. PDA takes advantage of {the fact that solutions consistent with the known dynamics will lie along a lower dimensional manifold in the higher dimensional sequence space}. A brief introduction of the PDA approach is given in the following paragraph (see~\cite{Judd2001,DuPDA} for additional details). 

{Given a dynamical system of dimension $m$ and a sequence of $n$ observations \footnote{For the Logistic Map, observations and system states share the same one dimensional space.}, define a sequence space as the $m\times n$ dimensional space in which a single point can be thought of as a particular series of $n$ states $\textbf{u}_{i}$, for $i=0,\dots,n-1$ where $\textbf{u}_{i}$ is an m dimensional vector. Most points in sequence space do not correspond to a trajectory of the system. Define a {\it pseudo-orbit}, $\textbf{U}\equiv \{\textbf{u}_{0},...,\textbf{u}_{n-2},\textbf{u}_{n-1}\}$, to be a point in the $m\times n$ dimensional sequence space for which $\textbf{u}_{i+1}\neq f(\textbf{u}_{i})$ for one or more components of $\textbf{U}$. Thus a pseudo-orbit corresponds to a sequence of states which is {\bf not} a trajectory of the system. Each sequence of $n$ observations $\textbf{s}_i, i=0,\dots,n-1$ defines a pseudo-orbit (a point in the sequence space). Call this an {\it observed pseudo-orbit}, $\textbf{S}\equiv \{\textbf{s}_{0},...,\textbf{s}_{n-2},\textbf{s}_{n-1}\}$, which with probability one will not be a trajectory. }

Define the {\it mismatch} to be: 
\begin{eqnarray}
 \label{eq:mismatch}
    e_{i}=\mid f(\textbf{u}_{i})-\textbf{u}_{i+1} \mid
\end{eqnarray}
By construction, trajectories have a mismatch of zero. The mismatch cost function is then given by:
\begin{eqnarray}
 \label{eq:miscost}
    C(\textbf{U})=\sum_{i=0}^{n-1} e_{i}^{2}
\end{eqnarray}
{The most common form of Pseudo-orbit Data Assimilation simply }minimizes the mismatch cost function for $\textbf{U}$ in the $m\times n$ dimensional sequence space {starting from the observed pseudo-orbit}. If a gradient descent (GD) approach is adopted\footnote{Other methods for this minimization are available, GD is discussed here due to its simplicity and adequacy.}, then {\bf a} minimum of the mismatch cost function can be obtained by solving the ordinary differential equation
\begin{eqnarray}
 \label{eq:odegd}
 \frac{d\textbf{U}}{d\tau}=-\nabla C(\textbf{U}),
\end{eqnarray}
where $\tau$ denotes algorithmic time.\footnote{The approach can be generalized to situations where gradient is not known analytically~\cite{2004JuddB}; improving the ability to work without gradient information would widen the application of the approach significantly.}
In practice, the minimization is initialized with the observed pseudo-orbit, i.e. $^{0}\textbf{U}=\textbf{S}$ where the pre-super-script {zero} on $\textbf{U}$ denotes the initial {algorithmic time} {($\tau=0$)} of the GD. {Under gradient descent, a point in sequence space moves towards the {trajectory }manifold under Equation~\ref{eq:odegd}}. { The minimization algorithm requires differentiating the mismatch cost function (Equation~\ref{eq:miscost}), which gives:
\begin{eqnarray}
\label{eq:pdcost}
\frac{\partial C(\textbf{U})}{\partial \textbf{u}_{i}}=2\times \left\{\begin{array}{ll}
                                                                  -(\textbf{u}_{i+1}-f(\textbf{u}_{i}))J(\textbf{u}_{i}) & i=0 \\
                                                                  -(\textbf{u}_{i}-f(\textbf{u}_{i-1}))+(\textbf{u}_{i+1}-f(\textbf{u}_{i}))J(\textbf{u}_{i}) & 0 < i < n-1  \\
                                                                  -\textbf{u}_{i}-f(\textbf{u}_{i-1}) & i=n-1
                                                                  \end{array}
                                                           \right.
\end{eqnarray}
where $J(\textbf{u}_{i})$ is the Jacobian of $f$ at $\textbf{u}_{i}$. The ordinary differential Equation~\ref{eq:odegd} is solved using the Euler approximation in the examples below.

}
The mismatch cost function has no local minima other than {points} on the manifold, for which $C(\textbf{U})=0$, ({this defines} the trajectory manifold\footnote{{Back-substitution of the solution of $-\textbf{u}_{n-1}-f(\textbf{u}_{n-2})=0$ into Equation~\ref{eq:pdcost} shows that the only critical points for $C(\textbf{U})$ have $\textbf{u}_{i}-f(\textbf{u}_{i-1})=0$ for all $i$ in $0 \leq i \leq n-1$. }All points on the trajectory manifold have zero mismatch (are trajectories) and only points on the trajectory manifold have zero mismatch.}) and {of course} every segment of trajectory lies on this manifold~\cite{Judd2001}. {Denote} the result of the GD minimization at {algorithmic time {$\tau=\alpha$ as} $^{\alpha}\textbf{U}\equiv ^{\alpha}\textbf{u}_{0},...,^{\alpha}\textbf{u}_{n-1}$}. Here $\alpha$ indicates {\it algorithmic time} in GD (i.e. the number of iterations of the GD minimization). As $\alpha \rightarrow \infty$, the pseudo-orbit $^{\alpha}\textbf{U}$ approaches a trajectory of the model asymptotically. In other words, the GD minimization takes the observed pseudo-orbit towards a model trajectory ($^{\infty}\textbf{U}$ is a point in sequence space on the trajectory manifold). 

In practice, the GD algorithm is run for a finite time and a trajectory is not obtained. {Nevertheless, the values of $^{\alpha}u_{0}$ for large $\alpha$, define candidates trajectories using information from the observation window $0\le i \le n-1$}. For $i>0$, the $i$-step preimages of the relevant component of $^{\alpha}u_{i}$ provide candidates for the initial state.
Complications arise from the fact that the Logistic Map is two-to-one; these {complications} have nothing to do with chaos per se\footnote{Beyond the fact that one-to-one maps in one-dimension cannot display chaotic dynamics, of course.}. {In }{one-to-one chaotic maps} the calculation of preimages {is }straightforward. 
The Logistic Map is a two-to-one map, and in most cases\footnote{Not in all cases, however. For discussion of the point see~\cite{Lalley99}.} only one of the two preimages for each $^{\alpha}u_{i}$ is relevant to $\tilde{x}_{i-1}$. In practice a threshold criteria to discard irrelevant preimages must be defined, a simple example would be to discard (with high probability) those preimages whose distance from the corresponding previous observation exceeds some threshold based on the {properties} of the observational noise (a $3\sigma$ criteria is {adopted} in the following section and shown to be adequate for purpose).

\subsection{Results}
\begin{figure}[h]
\centering
    \epsfig{file=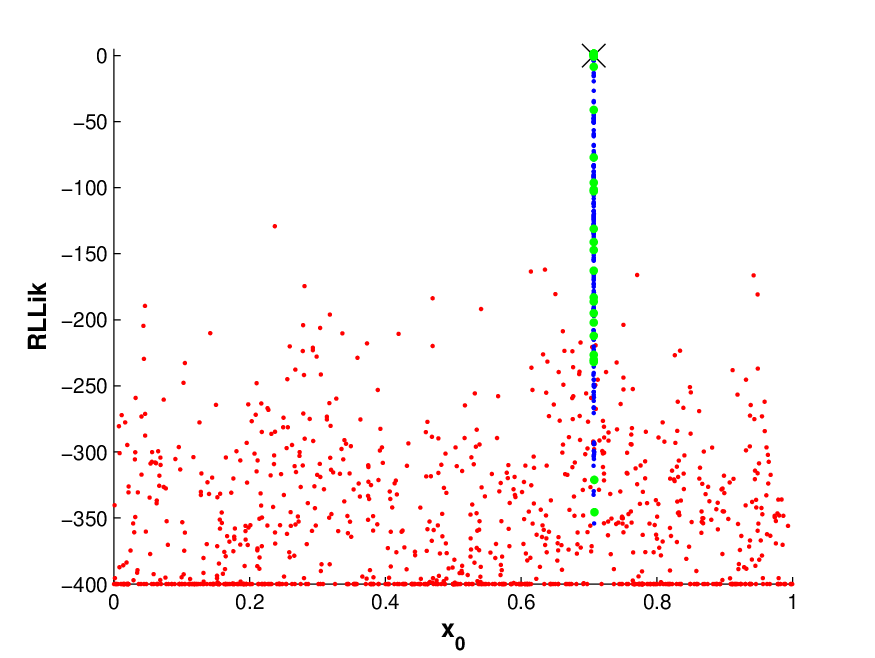, width=7cm}
    \epsfig{file=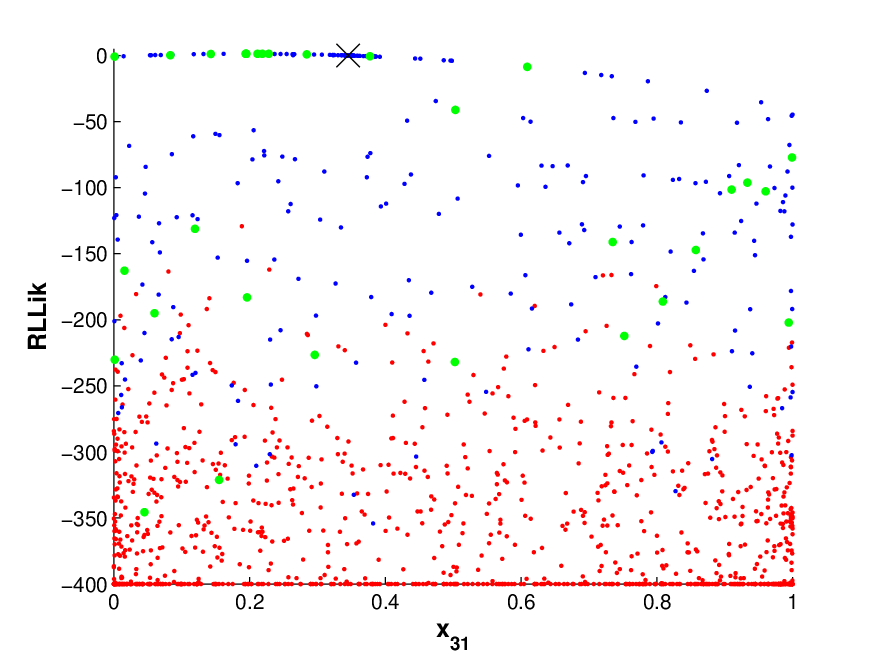, width=7cm}

\caption{{a) Following Figure~\ref{fig:LLikBlue}a, but in this case including he states located by PDA importance sampler are plotted in green. b) The forward images of those states plotted in (a) at time $31$.}}
  \label{fig:LLikGreen}
\end{figure}

The green points in Figure~\ref{fig:LLikGreen} {were} located using {PDA importance sampler; {note} that some have relative log-likelihood near zero (maximum 1.3)}. As expected, the observations do not contain sufficient information to identify the state of the system at the time of the final observation with the same degree of precision { as the initial state}. This is reflected in the fact that the green points are much less close to the true state at time $31$ (Figure~\ref{fig:LLikGreen}b) than {the corresponding green points} at time $0$ (Figure~\ref{fig:LLikGreen}a).

Two experiments were conducted to test the robustness of the PDA importance sampling approach. The first is based on $2048$ different realizations of observations given the same initial condition, $\tilde{x}_{0}=\sqrt{2}/2$, to examine consistency. The second considers $2048$ different initial conditions to examine robustness {over different states}. Three different observation window lengths were used in each experiment. Table~\ref{tab:LLikO} and Table~\ref{tab:LLikS} shows the results. {States with greater likelihood than the true state are often identified.}

Given uncertain observations, one can never identify Truth of a chaotic system unambiguously{; this} was noted {by Berliner $\&$ MacEachen~\cite{Berliner93,MacEachern95}, then Lalley~\cite{Lalley99,Lalley00}} and later explored by Judd $\&$ Smith~\cite{Judd2001}. Using the {PDA} approach, high likelihood states are indeed found: {states with} relative log-likelihood larger than minus one are found in every single experimental run. The fact that some PDA states have greater likelihood than Truth {reflects the fact} that Truth is not expected to be the most likely model state given the observations. 

For each experimental run, the minimum distance between those states obtained by PDA importance sampler (whose relative log-likelihood larger than minus one) and Truth is recorded. The minimum, maximum and median statistical values of the minimum distance from Truth are reported in Table~\ref{tab:LLikO} and \ref{tab:LLikS}. It is clear that the PDA importance sampler identifies states at higher quality (the minimum distance from Truth decreases) as the observation window length increases. This is expected inasmuch as more information from the dynamics {is} available {given} a longer window. Table~\ref{tab:LLikO} shows that the maximum value of the minimum distance among the $2048$ different realizations is $1.49\times 10^{-10}$ for a window length of 32 and in Table~\ref{tab:LLikS} the maximum value of the minimum distance among different true initial conditions is $2.02\times 10^{-10}$. The PDA importance sampler is both robust and efficient in this case.\footnote{Drawing samples uniformly from within a distance of $0.1$ of Truth would require $\sim 10^8$ candidates in order to find a candidate within $\sim 2\times 10^{-10}$ of Truth{; even then that close candidate need not have high likelihood.} The results of Table~\ref{tab:LLikO} and \ref{tab:LLikS} were obtained with only 1024 GD minimization iterations in each realization (each and every one of which identified high likelihood states close to Truth).} 


\begin{table}[th]
\centering

\begin{tabular}{|c|c|c|c|c|c|c|c|c|c|}
\hline
Window & \multicolumn{3}{|c|}{\texttt{\# of RLLik$>-1$}}  & \multicolumn{3}{|c|}{\texttt{\# of RLLik$>0$}} & \multicolumn{3}{|c|}{\texttt{Minimum distance to $\tilde{x}_{0}$}} \\
\cline{2-10}
length & \texttt{Min} & \texttt{Max} & \texttt{Median} & \texttt{Min} & \texttt{Max} & \texttt{Median} & \texttt{Min} & \texttt{Max} & \texttt{Median}\\
\hline

32 &  6 & 15  & 8  & 0 & 14 & 6 & $2.00\times 10^{-15}$ & $1.49\times 10^{-10}$ & $9.98\times 10^{-12}$\\

\hline

16 &  2 & 11  & 8  & 0 & 11 & 6 & $1.57\times 10^{-10}$ & $7.63\times 10^{-6}$ & $5.13\times 10^{-7}$\\

\hline

8 &  2 & 7  & 7  & 0 & 7 & 5 & $8.58\times 10^{-8}$ & $4.55\times 10^{-2}$ & $2.56\times 10^{-4}$\\

\hline



\end{tabular}

\caption{ Statistics of high likelihood states located by PDA importance sampler based on 2048 different realizations of observations (of $\tilde{x}_{0}=\sqrt{2}/2$) for the Logistic Map, i) statistics of the number of states (whose log-likelihood relative to the true state larger than minus one) ii) statistics of the number of states (whose relative log-likelihood larger than zero) iii) statistics of the minimum distance between the states (whose relative log-likelihood larger than minus one) and Truth.}
\label{tab:LLikO}
\end{table}

\begin{table}[th]
\centering

\begin{tabular}{|c|c|c|c|c|c|c|c|c|c|}
\hline
Window & \multicolumn{3}{|c|}{\texttt{\# of RLLik$>-1$}}  & \multicolumn{3}{|c|}{\texttt{\# of RLLik$>0$}} & \multicolumn{3}{|c|}{\texttt{Minimum distance to $\tilde{x}_{0}$}} \\
\cline{2-10}
length & \texttt{Min} & \texttt{Max} & \texttt{Median} & \texttt{Min} & \texttt{Max} & \texttt{Median} & \texttt{Min} & \texttt{Max} & \texttt{Median}\\
\hline

32 &  6 & 360  & 28  & 0 & 338 & 16 & $4.77\times 10^{-15}$ & $2.02\times 10^{-10}$ & $1.50\times 10^{-11}$\\

\hline

16 &  3 & 56  & 14  & 0 & 48 & 9 & $3.04\times 10^{-10}$ & $1.54\times 10^{-5}$ & $9.93\times 10^{-7}$\\

\hline

8 &  2 & 20  & 7  & 0 & 20 & 7 & $6.36\times 10^{-8}$ & $5.60\times 10^{-3}$ & $3.32\times 10^{-4}$\\

\hline



\end{tabular}

\caption{Statistics of high likelihood states located by PDA importance sampler based on 2048 different true initial states for the Logistic Map, i) statistics of the number of states (whose log-likelihood relative to the true state larger than minus one) ii) statistics of the number of states (whose relative log-likelihood larger than zero) iii) statistics of the minimum distance between the states (whose relative log-likelihood larger than minus one) and Truth.}
\label{tab:LLikS}
\end{table}

The experiments above {solve {by demonstration} Berliner's ``impossible" challenge; they} demonstrate that truly high likelihood points can be located using dynamical information, {easing} Berliner's identification problem of initial condition. Selecting an ensemble from this high likelihood set allows for informative forecasts which do not become useless until after those from the point forecasts illustrated by Berliner~\cite{Berliner1991} become uninformative.



\subsection{Equifinality and Relative Likelihood}
\begin{figure}[h]
\centering
  \epsfig{file=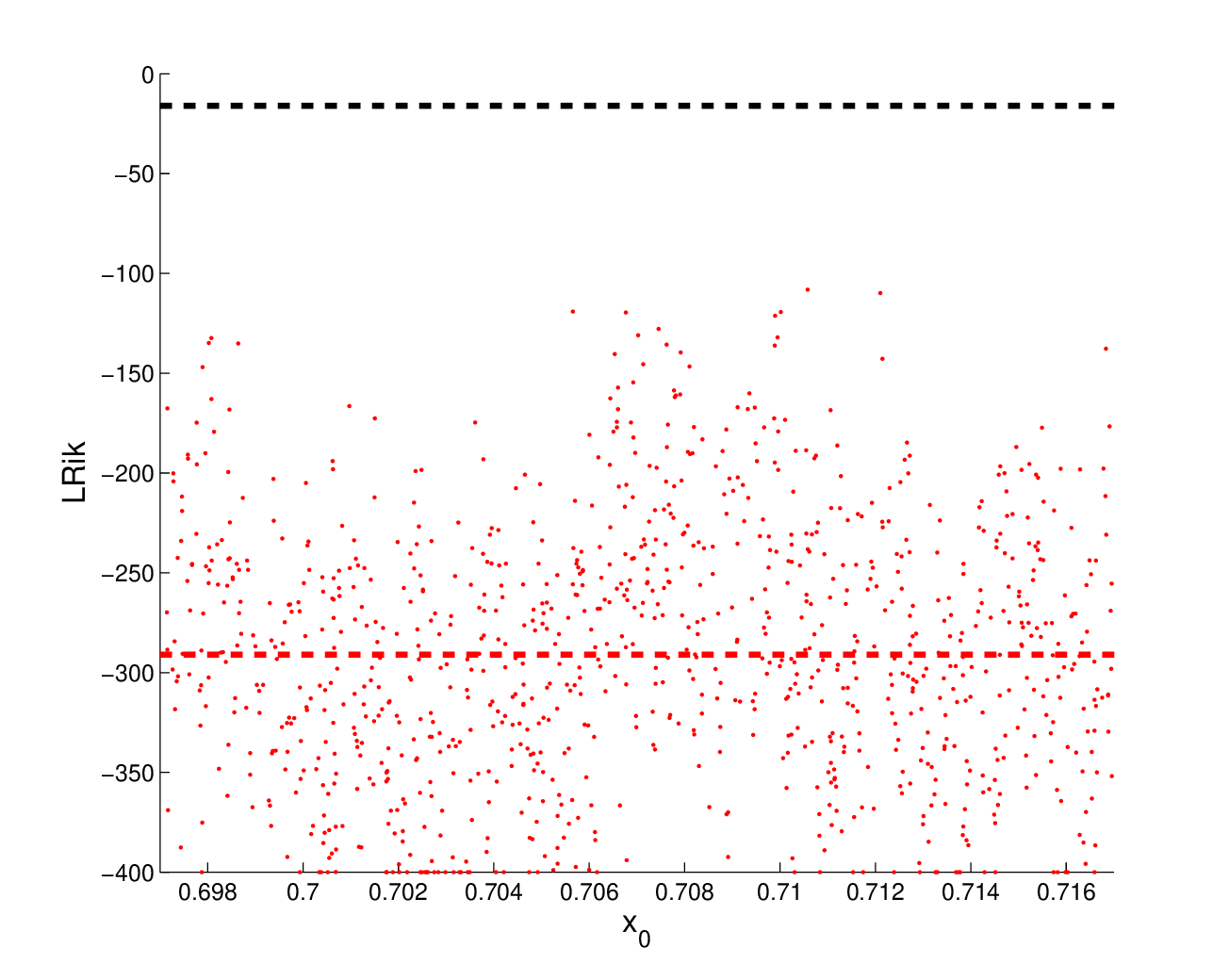, width=10cm}

\caption{Log-likelihoods of 1024 states uniformly sampled from  $[\tilde{x}_{0}-\frac{\sigma}{10},\tilde{x}_{0}+\frac{\sigma}{10}]$, the red dashed line is the median log-likelihood of those states and the black dashed line is the expected log-likelihood of Truth, following Figure~\ref{fig:LLikZoom}b.}
  \label{fig:LLikRef}
\end{figure}

Maximum likelihood estimation has been widely used~\cite{Pawitan01} since {its introduction} by Fisher~\cite{Fisher1922} in 1922. The ``best'' estimate is often chosen from a set of candidates, and only the relative likelihood of candidates within that sample is considered. Figure~\ref{fig:LLikRef} shows the log-likelihood of 1024 states (the same set used in Figure~\ref{fig:LLikZoom}b), the red dashed line is the median log-likelihood of those states. In this case, {the problem is not one of which estimate to select (equifinality), but one of recognizing that each and every candidate has vanishing likelihood (equidismality).} In practice, Truth is unknown therefore it cannot be used as a reference, {as it is in }Figure~\ref{fig:LLikZoom}. Given the observations and the noise model, however, the expected log-likelihood of Truth can be {determined}\footnote{The log-likelihood of Truth is $-\frac{\sum_{i=0}^{n-1}\delta_{i}^{2}}{2\sigma^{2}}$ (from Equation~\ref{eq:LLik}) where $\delta_{i}$ (observation noise) is $IID\sim N(0,\sigma^{2})$ distributed. Let $Z=\frac{\sum_{i=0}^{n-1}\delta_{i}^{2}}{\sigma^{2}}$, $Z$ is a random variable following chi-squared distribution with $n$ degrees of freedom. Statistics of the log-likelihood of Truth can therefore simply derived from $Z$.} and serve as a reference. Figure~\ref{fig:LLikRef}, the log-likelihood of 1024 states are plotted along with the expected log-likelihood of Truth (black dashed line). Figure~\ref{fig:LLikRef} shows that it is not a case of equifinality but a case of equidismality. In cases where it is observed that all traditional candidate states have vanishingly small log-likelihood relative to the expected log-likelihood of Truth, {a PDA based importance sampling might prove of valuable.}

{

\section{Implications for Quantifying Uncertainty in Practice}

{Having solved one of Berliner's ``impossible" challenges for nonlinear systems, a reviewer asks for potential implications this solution might hold in practice.} Two possibilities stand out. {The first regards improved noise reduction on the initial condition of the forecast via a PDA importance sampler; this would indeed be expected to improve forecast skill. Berliner's problem, however, focuses on identification of the point at the starting (oldest) observation, while weather forecasts are launched near the time of the end (most recent) observation. In operational weather forecasting, a relatively small set of initial conditions ($\sim 51$) are launched as forecasts in a relatively high dimensional space ($\sim 10^7$) in a Monte Carlo fashion every forecast cycle (typically every 6 hours); see~\cite{Leutbecher08,Hoskins13,Toth93}. As is long held in weather forecasting (see~\cite{Hansen01} for a discussion), uncertainty in the state at the start of the observation window in time collapses much more dramatically than that in the state at the end (present time); {this is consistent with results shown in Figure 4}. Figure 2 of~\cite{Hansen01} captures challenge Berliner $\&$ MacEachen~\cite{Berliner93,MacEachern95} , Hansen $\&$ Smith~\cite{Hansen01} and Lalley~\cite{Lalley00} foresaw. High likelihood states at present time lie along the unstable manifold, such sets of states are more efficient and informative, in terms of exploring initial condition uncertainty, than traditional approaches like Singular Vectors~\cite{Palmer94} or Bred Vectors~\cite{Toth97}. 

A second possibility lies in reanalysis~\cite{Parker15}. A reanalysis uses the most modern weather model to revisit and improve estimate of the state of the atmosphere of the distant past; here the opportunity for PDA importance sampler to make a contribution appears greater. A reanalysis takes a modern model and reconsiders past periods of time; reanalysis now common in atmospheric (weather) and oceanic contexts~\cite{Uppala05,Kalnay96,Oke08}. This task closely resembles that posed by Berliner. The results achieved in Table 1 and Table 2 reflect the aims of reanalysis directly, abet in a perfectly known one-dimensional system. Interpretations in terms of scalar ``noise reduction" are limited, as there can be information gained simply by restricting the state estimate to the model manifold. 

Reanalysis have many applications, one of significant current interest is index based insurance. In this case, uncertainty quantification of the index would be of significant interest. In short, the complications of evaluating insurance claims based on certifying damages in far flung regions of the world has lead to the insurance policies triggered by a model based meteorological index.~\cite{CCRIF15} 
To the extent that the methods deployed in this paper better quantify (and reduce) the uncertainty in the wind speeds of a storm within a few days of its passing, the efficiency of index based insurance could be improved.

Note from Table 1 and 2 that the longer the window length the better estimate (noise reduction) is achieved. Noise reduction based upon PDA importance sampler might also help fill in missing historical observations, as well identify model deficiency when identifying model trajectories intended to reflect past observations.}

\section{Distinguishing between initial conditions and parameters}
{The likelihood functions of initial conditions and that of parameter values have similar features~\cite{Berliner1991}. There are, however,} fundamental differences in the information available to address these two distinct estimation problems. 

Given the structure of the model class, the model parameter value determines dynamical behaviors of the model (e.g. natural measures) which are not changed by the initial condition. Given the model and {``true" parameter value, the search for unknown initial conditions is aided by the restriction the nature measure(s) places on candidate initial conditions in the state space (and, thereby, on trajectories in the sequence space). }

It is unclear how to construct similar constraints on {unknown} parameter values in the parameter space given the ``true'' initial state (if they exist\footnote{Neither is it clear how to construct the set of parameter values whose corresponding invariant measure contains the ``true'' initial state, or how to exploit this set; {what is clear is} how to exploit the existence of trajectory manifold given a particular value of the parameter.}). Uncertainty in initial state differs from uncertainty in the parameter value. The information in a {\it measurement} of the initial condition uncertainty will decay with time and eventually become statistically indistinguishable from a random sample of the natural measure, while the information {in the initial condition is preserved, arguably forever. }

While assuming the parameter value is perfect {might initially appear extreme}, it {appears less} nonsensical given that one has already assumed that the model structure is perfect. Assuming the initial state is perfect indicates a noise free observation is possible. Let the model's parameters be contained in the vector $\textbf{a}\in \mathbb{R}^{l}$. A set of $l+1$ sequential noise free observations $s_{i}, s_{i+1},...,s_{i+l}$ would, in general, be sufficient to determine $\textbf{a}$~\cite{McSharry1999}. {Thus if} one noise free observation is obtainable, obtaining only a few more noise free observations would define the true parameter value precisely. {Perhaps a} more realistic way to put the problem is to estimate the parameter value(s) given {realistic} observations, without assuming the ``true'' initial condition is known. In that case, the goal is to locate high likelihood trajectories (Smith et al.~\cite{Smith2010} call these shadowing trajectories) {for particular sets of} parameter values. {It is not yet} clear how to solve such a problem. In fact, it is not clear {if it is possible} to constrain the solution in the parameter space in a manner {analogous to} the constraints in the space of initial condition achieved by using trajectory manifold in the state space. 

Given a perfect model structure and knowing the true parameter value(s), the true initial state is a well defined goal of the identification. Inasmuch as structural model errors imply no true parameter value exists~\cite{Judd2004,DuPDA2}, it is unclear how one might define ``true'' initial state and the goal of estimation must be rethought when the structure of the system differs from that of the model.

Despite the importance of model parameters, {outside linear systems there is no general method of parameter estimation \footnote{{That is not to dismiss important advances in parameter estimation when attention is restricted to certain particular classes of model structure~\cite{Cressie11}; our claim here is that no general approach to parameter estimation is known which is applicable to nonlinear models in general}.}.} Methods have been developed to obtain useful parameter values with some success: McSharry and Smith~\cite{McSharry1999,Sornette} estimate model parameters by incorporating the global behaviour of the model into the selection criteria; Creveling et al.~\cite{Creveling2008} have exploited synchronization for parameter estimation; Smith et al.~\cite{Smith2010} focused on the geometric properties of trajectories; Du and Smith~\cite{Du2011} select parameter values based on the Ignorance score of ensemble forecasts. Each of these methods, however, require a large set of observations. Challenges remain when only a short sequence of observations is available. {Berliner's second challenge still stands.}

\section{Conclusion}

Berliner illustrated that even in the perfect model scenario traditional approaches are unable to provide good estimates of the initial condition for nonlinear chaotic systems. In large part, {those failures are} due to the inability of traditional approaches to skillfully meld the information in the dynamics of the nonlinear system itself with that in the observations. The importance sampling approach presented here {uses Pseudo-orbit Data Assimilation to} combine information from both observations and dynamics more effectively, thereby locating high likelihood initial states; this achieves {one} aim Berliner (1991) argued to be impossible. {As discussed in Section 5, this achievement may prove useful in reanalysis studies.} 

Despite the similarity of state estimation and parameter estimation, there are fundamental differences between uncertainty in the initial state and uncertainty in parameter value. Significant obstacles remain in solving {Berliner's second} challenge regarding parameter estimation while Pseudo-orbit Data Assimilation overcomes his first challenge regarding initial condition. 

\section*{Acknowledgment}
This research was supported by the LSE's Grantham Research Institute on Climate Change and the Environment and the ESRC Centre for Climate Change Economics and Policy, funded by the Economic and Social Research Council and Munich Re. Additional support for H.D. was also provided by the National Science Foundation Award No. 0951576 ``DMUU: Center for Robust Decision Making on Climate and Energy Policy (RDCEP)". L.A.S. gratefully acknowledges the continuing support of Pembroke College, Oxford. We also acknowledge the insightful comments of Michael Stein and Mark Berliner in discussions of this work.

\end{document}